\documentclass[amsmath,amssymb,twocolumn]{revtex4}
\usepackage[parfill]{parskip}    
\usepackage{graphicx}
\usepackage{amssymb}
\usepackage{epstopdf}
\usepackage{color}
\usepackage{graphicx}
\usepackage{pdfpages}

\begin{document}

\title{Scale--dependent polytropic black hole}
\author{Ernesto Contreras {${}^{a}$} \footnote{On leave 
from Universidad Central de Venezuela}
\footnote{ej.contreras@uniandes.edu.co} Angel Rincon {${}^{b}$} Benjamin Koch {${}^{b}$}
Pedro Bargue\~no{${}^{a}$}\footnote{p.bargueno@uniandes.edu.co}}
\address{${}^a$Departamento de F\'{\i}sica,
Universidad de los Andes, Apartado A\'ereo {\it 4976}, Bogot\'a, Distrito Capital, Colombia\\
${}^b$Instituto de F{\'i}sica, Pontificia Universidad Cat{\'o}lica de Chile,\\ Av. Vicu{\~n}a Mackenna 4860, Santiago, Chile.}

\begin{abstract}
In the present work we study the scale--dependence of polytropic non-charged black holes
in (3+1)-dimensional space--times assuming a cosmological constant. We allow 
for scale--dependence of the gravitational and cosmological couplings, and we solve the 
corresponding generalized field equations 
imposing the null energy condition. Besides, some properties, such as horizon structure and thermodynamics, are discussed in detail.
\end{abstract}

\maketitle

\section{Introduction}\label{intro}
The polytropic equations of state are appropriate in many situations in the context of General Relativity as well as in astrophysical problems.
For example, it is well known that astrophysical objects as cores of stars, fully convective low-mass stars, white 
dwarfs, neutron stars and galactic halos can be modelled with matter which fullfills a polytropic equation 
of state \cite{tooper1964,tooper1965,tooper1966,bludman1973,nilsson2000,maeda2002,herrera2004,lai2009,thirukkanesh2012,
Stuchlik2016,Stuchlik2017}.
In addition, polytropic equations of state have been considered in a cosmological context in order to model the matter
content \cite{mukhopadhyay2008}.

Recently, black hole (hereafter BH) solutions have been obtained considering polytropic equations 
of state \cite{setare2015}. In that work, the authors map the negative cosmological coupling with an effective pressure,
demanding that it obeys a polytropic equation of state. After that, the matter content degrees of freedom are eliminated from the Einstein 
field equations and, finally, solutions matching polytropic thermodynamics with that of BHs are obtained. \\
The aforementioned solutions are obtained in the context of classical gravity. However, it is very well known that 
a more complete description quantum
effects must be considered. As the full theory of quantum gravity is still lacking,
many different works have been devoted to get some insight into the underlying physics (for an incomplete list check 
\cite{Wheeler:1957mu,Deser:1976eh,Rovelli:1997yv,Bombelli:1987aa,Ashtekar:2004vs,Sakharov:1967pk,Jacobson:1995ab,Verlinde:2010hp,Reuter:1996cp,Litim:2003vp,Horava:2009uw,Charmousis:2009tc,Ashtekar:1981sf,Penrose:1986ca,Connes:1996gi,Nicolini:2008aj,Gambini:2004vz} and for a review see \cite{Kiefer:2005uk}). Despite the fact that in those works the authors discuss different aspects of 
quantum gravity, most of them have the common feature that the resulting effective gravitational action acquires a scale--dependence. 
This behaviour is observed through the couplings of the effective action: they change from fixed values to scale--dependent quantities, 
i.e. $\{G_0, \Lambda_0\} \mapsto \{G_k, \Lambda_k\}$, where $G_0$ is Newton's coupling and $\Lambda_0$ is the cosmological coupling. Indeed, there 
is some evidence which supports that this scaling behaviour is consistent with Weinberg's Asymptotic Safety program 
\cite{Weinberg:1979,Wetterich:1992yh,Dou:1997fg,Souma:1999at,Reuter:2001ag,Fischer:2006fz,Percacci:2007sz,Litim:2008tt}. 
In addition, the effective action assuming running couplings has been studied in three--dimensional space--times in 
the context of BH physics in Ref. \cite{Koch:2014joa, koch2016, Rincon:2017goj, Rincon:2017ypd, Rincon:2017ayr,Contreras:2017eza}, 
in four dimensions \cite{Koch:2014joab}
and in the cosmological context 
\cite{Hernandez-Arboleda:2018qdo}.
In the aforementioned works, the corresponding scale--dependent couplings take into account a quantum effect, 
namely, this approach admits corrections to both the classical BH backgrounds and to the FLRW universe. 

Then, inspired by this fact, the next step is to take advantage of the aforementioned approach to produce interesting BH
solutions in new scenarios. The Van der Waals black hole \cite{Rajagopal:2014ewa} is a recent solution which, after 
identifying $P$ with the cosmological constant $\Lambda$, allow us to write down a BH equation of state $P=P(V, T )$ and 
to compare it to the corresponding fluid equation of state \cite{Rajagopal:2014ewa}. We remark that this fact opened a window to 
investigate the analogy between BHs and certain fluids (for a recent review, see \cite{kubiznak2017}). 

In this work, 
inspired by this idea, we study how a polytropic BH in four dimensional space--time obtained in Ref. \cite{setare2015} is 
generalized to the case of scale--dependent couplings,
 implemented and set at the level of an effective action.

The work is organized as follows: In Sect. \ref{pol_theory} we introduce and summarize the polytropic 
BH solution, whereas Sect. \ref{scale_theory} is devoted
to briefly introduce the scale--dependent gravitational setting, which is employed in Sect. \ref{poly_BH_section} to obtain the
scale--dependent BH solution. Along this section, the new solution is carefully studied with emphasys on horizons, asymptotics, singularities, thermodynamics
and their comparison with the previously studied 
classical polytropic solution. Finally, some concluding remarks are given in Sect. \ref{remarks}.

\section{Polytropic Black Hole Solution}\label{pol_theory}
In this section we sumarize the main results obtained in Ref. \cite{setare2015}. 
The line element is parametrized as
\begin{align}\label{metric}
&\mathrm{d} s^{2}=-f_{0}(r) \mathrm{d} t^{2}+f_{0}(r)^{-1} \mathrm{d}r^{2}+r^{2}\mathrm{d}\Omega^{2},
\end{align}
where the lapse function is computed to be
\begin{align}
&f_{0}(r,P_{0})=\frac{r^{2}}{L^{2}}-\frac{2G_{0}M_{0}}{r}+h(r,P_{0}).
\end{align}
On one hand, the parameter $P_0$ is associated with a negative cosmological constant, $\Lambda_0$,
and with the parameter $L^{2}$ by
\begin{eqnarray}\label{preslamb}
P_{0}:=-\frac{\Lambda_{0}}{8\pi G_{0}}=\frac{3}{8\pi G_{0} L^{2}} =\frac{3}{\kappa_0 L^2}.
\end{eqnarray}
In addition, note that $\kappa_0$ is the so--called Einstein constant, which is defined in terms of the Newton coupling, $\kappa_0 \equiv 8\pi G_0$. On the other
hand, $P_{0}$ is taken as the pressure of a polytropic gas with equation of state
\begin{eqnarray}\label{depes}
P_{0}=K\varrho^{1+\frac{1}{n}}, 
\end{eqnarray}
where $\varrho$ is the polytropic gas density, $K$ is a positive constant and $n$ is the polytropic 
index which takes values $0\leq n<\infty$ \cite{Chandrasekhar:1957}.

In order to obtain the the unknown function $h(r,P_{0})$, the authors of Ref. \cite{setare2015} assume that
\begin{itemize}
 \item The metric (\ref{metric}) is a solution of the Einstein field
 equations $G_{\mu\nu}+g_{\mu\nu}\Lambda_{0} = \kappa_0 T_{\mu\nu}$ with $\Lambda_0<0$ and
$T^{\mu}_{\nu}=\nobreak diag(-\varrho,p_{1},p_{2},p_{3})$.
 \item The thermodynamics of the BH solution is matched with that of a polytropic gas after eliminating 
 the matter degrees of freedom from the Einstein field equations.
\end{itemize}
With these assumptions, $h(r,P_{0})$ reads
\begin{align}\label{unk}
&h = C_1 r^{\frac{1-n}{1+n}} \left(K^{-\frac{n}{n+1}}+P_{0}^{\frac{1}{n+1}}\right)+\frac{C_2}{r}+\frac{1}{3} \kappa_0 P_{0} r^2.
\end{align}
In particular, for $n=-\frac{1}{3}$, $C_{2}=0$ and
\begin{eqnarray}\label{incons}
C_{1}=-\frac{1}{3} \kappa_0 P_{0}\left(K^{-\frac{n}{1+n}}+P_{0}^{\frac{1}{1+n}}\right)^{-1}.
\end{eqnarray}
Therefore, one obtains
\begin{align}
&f_{0}=\frac{r^{2}}{L^{2}}-\frac{2G_{0}M_{0}}{r}, \label{metricdef}
\\
&\rho_{0}=-p_{1}=\frac{1}{\kappa_0 r^{2}}, \label{rhop}
\\
&p_{2}=p_{3}=0.\label{pdos}
\end{align}
Note that for this choice of the integration constants, the parameters $\rho_{0}$
and $p_{i}$ ($i=1,2,3$) describe a perfect fluid \footnote{Note that, although the pressures are non--equal, we refer to a
perfect fluid in the sense that $\rho$ and $p$ are related by
$\rho_{0}=-p_{1}$} as a matter--source of the Einstein field 
equations and the spacetime results asymptotically anti--de Sitter.

It is worth mentioning that, 
in this case, the matter content does not depend on the constant
$K$ appearing in the polytropic equation of state for the dark energy content. However, this is not true in general because 
the result could be affected by the particular choice of the polytropic index $n$.

It is remarkable that $T_{\mu\nu}$ fullfils
all the energy conditions
\begin{align}
&\rho_{0}\ge0,          
\hspace*{2.79cm}
\rho_{0}+p_{i}\ge0,\\
&\rho_{0}+\sum\limits_{i}p_{i}\ge 0,
\hspace*{1.5cm}
\rho_{0}+p_{i}\ge0,\\
&\rho_{0}\ge|p_{i}|,
\end{align}
which are referred as the weak, strong and dominant energy conditions, respectively.

Besides, the knowledge of the invariants (scalars) allows to check if the theory presents some singularity. Thus, the corresponding 
scalars are computed below: 
\begin{align}
&R_{0}=2 \left(\frac{1}{r^2}-\frac{6}{L^2}\right), 
\\
&Ricc_{0}= 2 \left(\frac{18}{L^4}-\frac{6}{L^2 r^2}+\frac{1}{r^4}\right),
\\
&\mathcal{K}_{0}=\frac{4 \left(L^4 \left(12 M_{0}^2+4 M_{0} r+r^2\right)-2 L^2 r^4+6 r^6\right)}{L^4 r^6},
\end{align}
where $R_0$, $Ricc_{0}$ and $\mathcal{K}_{0}$ are the classical Ricci, Ricci squared and Kretschmann scalars, respectively. One 
observes that they
have a divergence at $r=0$. In this sense, the BH solution is singular.
In order to get insight into the thermodynamics properties of this
BH, we obtain the event horizon which is located at
\begin{align}
&r_{0}= \left(2G_{0}M_{0} L^2 \right)^{1/3}.
\end{align}
Rewriting the lapse function in terms of the event horizon we have 
\begin{eqnarray}
f_{0}(r)=\frac{1}{L^2 r}\left(r^3 - r_{0}^3\right).
\end{eqnarray}
The thermodynamic behaviour of the BH can be characterized by the corresponding Hawking temperature, $T_{0}$, and the Bekenstein--Hawking entropy, $S_{0}$, 
which are given by
\begin{align}
&T_{0} = \frac{1}{4\pi}\left|\lim_{r \rightarrow r_0}\frac{\partial_r g_{tt}}{\sqrt{-g_{tt} g_{rr}}}\right| = \frac{3}{4\pi}\left| \frac{r_{0} }{L^{2}}\right|,
\\
&S_{0} = \frac{\mathcal{A}_H(r_0)}{4 G_0}. 
\end{align}
Note that $\mathcal{A}_H$ is the horizon area which is given by
\begin{align}
&\mathcal{A}_H(r_0) \equiv \oint \mathrm{d}x^{2} \sqrt{h} =4 \pi r_0^2,
\end{align}
where $h_{ij}$ is the induced metric at the event horizon $r_0$.
After this brief summary of the polytropic BH, in the next section we will study the scale--dependent setting.

\section{Scale--dependent coupling and scale--setting}\label{scale_theory}

The scale--setting procedure presented in the first part of this section
follows closely the spirit and concept of Ref. \cite{Koch:2014joa}.
The scale--dependent effective action in the Einstein--Hilbert truncation reads
\begin{eqnarray}\label{action}
\Gamma[g_{\mu\nu},k]=\int \mathrm{d}^{4}x\sqrt{-g}\bigg[\frac{1}{2 \kappa_k} (R-2\Lambda_{k}) 
+\mathcal{L}^{\text{M}}_k\bigg],
\end{eqnarray}
where $G_{k}$ and $\Lambda_{k}$ stand for the scale--dependent
 gravitational and cosmological coupling, respectively, whereas $\kappa_k \equiv 8 \pi G_k$ is the scale--dependent Einstein coupling
and $\mathcal{L}_{k}^{\text{M}}$ is the Lagrangian density for the matter content.\\
Variations with respect to the metric field $g_{\mu\nu}$,  
give
the modified Einstein field equations 
\begin{eqnarray}\label{einstein}
G_{\mu\nu}+g_{\mu\nu}\Lambda_{k}=8\pi G_{k}T^{eff}_{\mu\nu},
\end{eqnarray}
where
$T^{eff}_{\mu\nu}$ is the effective energy--momentum tensor defined as
\begin{eqnarray}\label{eff}
T^{eff}_{\mu\nu}:=(T^{\text{M}}_{k})_{\mu\nu} - \frac{1}{8\pi G_{k}}\Delta t_{\mu\nu}.
\end{eqnarray}
In Eq. (\ref{eff}), $(T^{\text{M}}_{k})_{\mu\nu}$ is the matter energy--momentum tensor and $\Delta t_{\mu\nu}$ is given by
\begin{eqnarray}\label{nme}
\Delta t_{\mu\nu}=G_{k}\left(g_{\mu\nu}\square -\nabla_{\mu}\nabla_{\nu}\right)G_{k}^{-1}.
\end{eqnarray}
Please note that the renormalization scale $k$ is, indeed, not constant anymore, which means 
that the stress energy tensor is likely not conserved, as was discussed previously in Ref. \cite{Rincon:2017goj}. This kind of problem 
has been considered in the context of renormalization group improvement of BHs in asymptotic safety 
scenarios (see, for instance \cite{Bonanno:2000ep,Bonanno:2006eu,Reuter:2010xb,Koch:2014cqa} and references therein).

This inconsistency with very fundamental conservation laws 
can be avoided by applying the variational scale--setting procedure described in Ref. \cite{Koch:2014joa},
where the metric equations of motion (\ref{einstein}) are complemented by
an equation obtained from 
variations with respect to the scale--field $k(x)$
\begin{align}\label{scale}
\frac{\mathrm{d}}{\mathrm{d} k} \Gamma[g_{\mu \nu}, k] =0.
\end{align}
It is worth mentioning that, 
if the precise beta functions of the problem are not known, the
 Eqs. (\ref{action}) and (\ref{scale}) do not have enough information in order 
to solve for the two independent fields, $g_{\mu\nu}(x)$ and $k(x)$.
In order to solve this problem, we take into
account the so--called null energy condition and assume that the couplings $\{G_k$, $\Lambda_k\}$ depend explicitly on space-time coordinates,
a dependence which is inherited from the space-time dependence of $k(x)$. 
Thus, as reported in Refs. \cite{Koch:2014joa,Koch:2014joab,koch2016,Rincon:2017goj,Rincon:2017ypd,
Hernandez-Arboleda:2018qdo}, 
we can encode the ignorance on the scale--dependence of the coupling parameters by promoting $G_{k}$ and $\Lambda_{k}$ to 
independent fields, $G(x)$ and $\Lambda(x)$, and considering Eq. (\ref{einstein}) with some extra asumption in order to solve for the unknown
functions.
\\
In this work we follow the approach presented in Refs. 
\cite{Koch:2014joa,Koch:2014joab,koch2016,Rincon:2017goj,Rincon:2017ypd,Hernandez-Arboleda:2018qdo}. Moreover, we assume a static
and spherically symmetric space--time and a line element parametrized as
\begin{eqnarray}\label{metricp}
ds^2=-f(r)dt^2 + f(r)^{-1} dr^2+r^2 d\Omega^2. 
\end{eqnarray}
Note that replacing Eq. (\ref{metricp}) in Eq. (\ref{einstein}) we shall obtain two independent differential equations for the three independent fields
$f(r)$, $G(r)$ and $\Lambda(r)$. An alternative way to decrease the number of degrees of freedom consists in demanding some energy condition on 
$T^{eff}_{\mu\nu}$ \cite{rubakov}. It is well known that the null energy condition
 is the least restrictive of the usual energy conditions and that it can help
to obtain suitable solutions of the Einstein field equations \cite{koch2016}. For $T^{eff}_{\mu\nu}$, the null energy
condition is
\begin{align}
T^{eff}_{\mu\nu}\ell^{\mu}\ell^{\nu} \geq 0,
\end{align}
which reads 
\begin{align}
\left[(T^{\mathrm{M}})_{\mu\nu} - \frac{1}{8\pi G(r)}\Delta t_{\mu\nu}\right]\ell^{\mu}\ell^{\nu} \ge 0,
\end{align}
where $\ell^{\mu}$ is a null vector. Note that $(T^{\text{M}})_{\mu\nu}$
corresponds to the matter stress--energy tensor after replacing
$G_{0}\rightarrow G(r)$.
Considering the special case $\ell^{\mu}=\{f^{-1/2},f^{1/2},0,0 \}$
we obtain that $(T^{\mathrm{M}})_{\mu\nu}\ell^{\mu}\ell^{\nu}=0$ which implies that $\Delta t_{\mu\nu}\ell^{\mu}\ell^{\nu}\ge 0$.
However, it can be shown that $G_{\mu\nu}\ell^{\mu}\ell^{\nu}=0$ and, thus by virtue of (\ref{einstein}), 
one finds
\begin{eqnarray}\label{necnm}
\Delta t_{\mu\nu}\ell^{\mu}\ell^{\nu}=0.
\end{eqnarray}
This equation encodes the radial dependence of Newton's coupling $G(r)$ \cite{Rincon:2017ayr}.
In particular, the corresponding differential equation is 
\begin{align}\label{EDO_G}
2\left[\frac{\mathrm{d} G(r)}{\mathrm{d} r}\right]^2 = G(r)\frac{\mathrm{d}^2 G(r) }{\mathrm{d} r^2 }.
\end{align}
Therefore, solving Eq. (\ref{EDO_G}) we obtain
\begin{eqnarray}\label{gr}
G(r)=\frac{G_{0}}{1+\epsilon r},
\end{eqnarray}
where $\epsilon\ge 0$ is an integration constant with dimensions of inverse of length. 
Note that in the limit $\epsilon\to 0$, $G(r)=G_{0}$, 
which implies that $\Delta t_{\mu\nu}=0$ and thus, the classical Einstein field equations are recovered. 
Therefore, the strength of scale--dependence
is controlled by the so called running parameter $\epsilon$.

\section{Scale--dependent polytropic black hole}\label{poly_BH_section}

In this section we obtain solutions for the modified Einstein field equations
\begin{eqnarray}\label{einsteinr}
G_{\mu\nu}-8\pi G(r)P(r)g_{\mu\nu} = 8\pi G(r)(T^{\text{M}})_{\mu\nu} -\Delta t_{\mu\nu}.
\end{eqnarray}
In the above expression, $G(r)$ is given by Eq. (\ref{gr}), $P(r)=-\Lambda(r)/8\pi G(r)$ stands 
for the scale--dependent polytropic pressure and 
\begin{eqnarray}\label{tmunumatt}
(T^{\text{M}})^{\mu}_{\nu}
= diag(-\tilde{\rho},\tilde{p},0,0) ,
\end{eqnarray}
with $\tilde{\rho}=-\tilde{p}=1/(8\pi G(r)r^2)$. 
%
Besides, it should be noted that the scale--dependent pressure is linked with the energy density
through a fixed number and not by some scale--dependent coupling. This is because we encode the scale--dependence into the pressure 
and the energy density, assumption which is always possible. More precisely, the parametrization of $\tilde{\rho}$ and $\tilde{p}$ can be
thought as a generalization of the classical results in Eq. (\ref{rhop}) after the incorporation of the scale-dependent coupling $G(r)$.

\subsection{Solutions}
\begin{figure}[ht!]
\centering
\includegraphics[width=\linewidth]{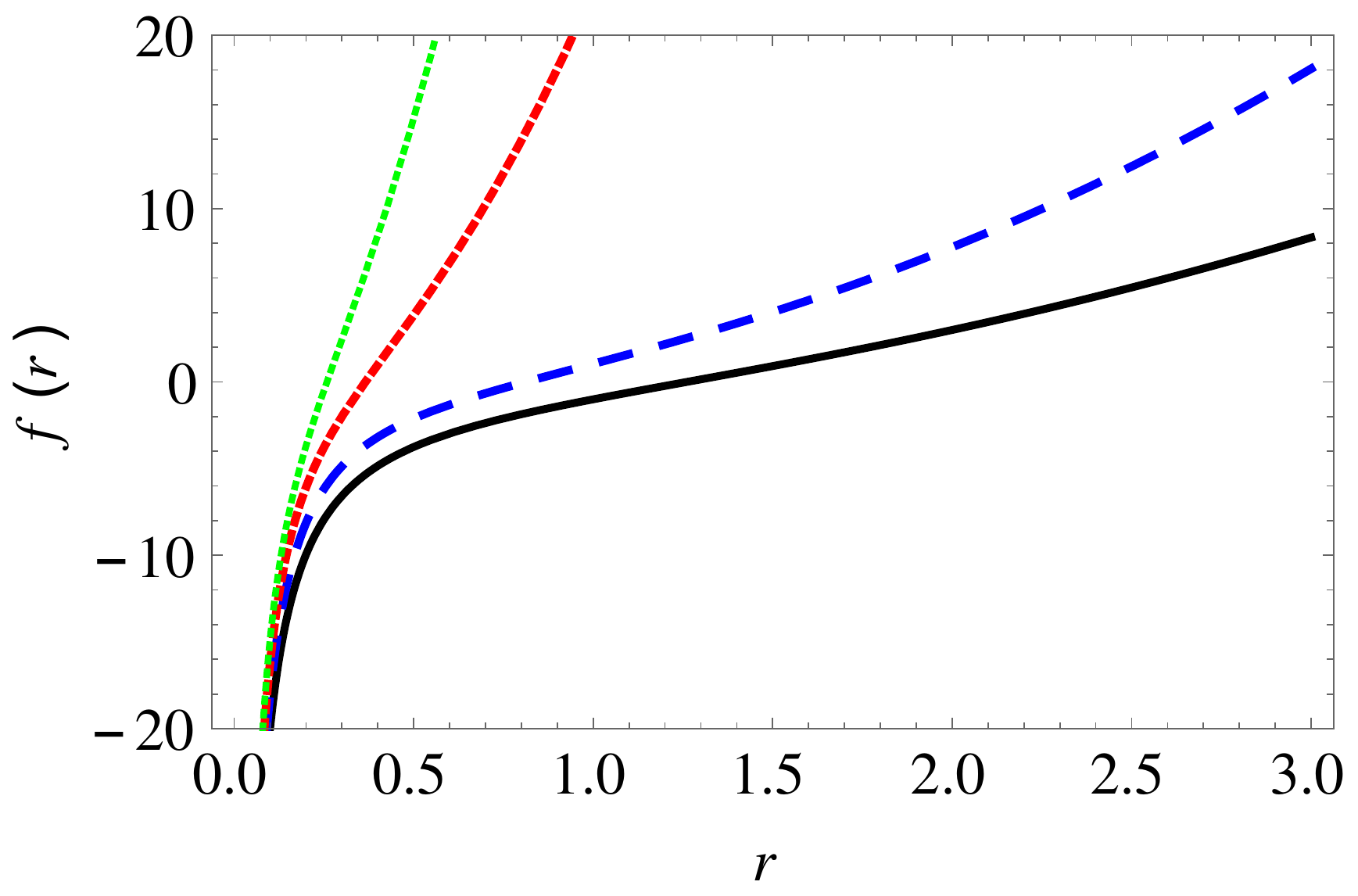}
\caption{\label{lapsefig} 
Lapse function for
$\epsilon=0.00$ (black solid line), $\epsilon=0.75$ (dashed blue line), $\epsilon=1.50$ 
(short dashed red line) and $\epsilon=2.00$ (dotted green line). The other values have been taken as unity. See text for details.
}
\end{figure}

The solution for the scale--dependent polytropic BH is given by
\begin{equation}
\begin{split}
&f(r) = f_{0}(r)+6 G_{0} M_{0} r^2 \epsilon ^3 \ln \left[2 G_{0} M_{0} \frac{r \epsilon +1}{r}\right] \\
& \hspace{1cm} +3G_{0}M_{0}\epsilon  (1-2 r \epsilon ), \label{lapsepoly}
\end{split}
\end{equation}
\begin{equation}
\begin{split}
&P(r) =(2 r \epsilon +1)P_{0}
-\frac{3 M_{0}}{8 \pi }\frac{ (12 r \epsilon  (r \epsilon +1)+1)\epsilon ^2}{ r (r \epsilon +1) } \\
&\hspace{1cm} +\frac{18 G_{0} M_{0}  \epsilon ^3 (2 r \epsilon +1)}{8 \pi  G_{0} } \ln \left[2 G_{0} M_{0}\frac{ r \epsilon +1}{r}\right],
\end{split}
\end{equation}
\begin{align}
&\tilde{\rho}(r) =-\tilde{p}=(1+\epsilon r)\rho_{0} =\frac{1+ \epsilon r}{8\pi G_{0} r^2},
\\
&G(r) = \frac{G_{0}}{1+\epsilon r},
\end{align} 
Note that our solution reproduces the classical results given by Eqs. (\ref{metricdef}), (\ref{rhop}) and (\ref{preslamb}) in the limit $\epsilon\to0$. 
In fact,
\begin{align}
&\lim\limits_{\epsilon\to0}f(r) = \frac{r^{2}}{L^{2}}-\frac{2G_{0}M_{0}}{r},\\
&\lim\limits_{\epsilon\to0}\tilde{\rho}(r) = \frac{1}{8\pi G_{0}r^2},\\
&\lim\limits_{\epsilon\to0}\tilde{p}(r) = -\frac{1}{8\pi G_{0}r^2},\\
&\lim\limits_{\epsilon\to0}P(r) = -\frac{\Lambda_0}{8\pi G_{0}}=\frac{3}{8\pi G_{0} L^{2}},
\end{align}
as expected.\\
In figure \ref{lapsefig} the lapse function is shown for different values
of $\epsilon$ compared to the classical BH solution,
showing that, for small $\epsilon$ values, the scale--dependent lapse function coincides with the classical one. 
However, when $\epsilon$
increases, a deviation from the classical solution appears. In particular, when $r\to\infty$, 
\begin{align}\label{asympf}
&f(r) \approx \frac{r^2}{L^2} +  6 G_{0} M_{0} r^2 \epsilon ^3 \ln (2 G_{0} M_{0} \epsilon).
\end{align}
We note that, asymptotically, the scale--dependent effect only dominates when high values of the running parameter $\epsilon$ are considered. In addition,
as Eq. (\ref{asympf}) shows, the scale--dependent lapse function behaves asymptotically as anti--de Sitter as in the classical case but
with an effective cosmological constant given by
\begin{eqnarray}
\Lambda_{eff}=-\left[\frac{1}{L^2}+6 G_{0} M_{0} \epsilon ^3 \ln (2 G_{0} M_{0} \epsilon )\right],  
\end{eqnarray}
provided  
\begin{eqnarray}
\epsilon > (2G_{0} M_{0})^{-1},
\end{eqnarray}
or 
\begin{align}
&\epsilon < (2G_{0} M_{0})^{-1},
\\
&\frac{1}{L^{2}} > -6 G_{0} M_{0} \epsilon ^3 \ln (2 G_{0} M_{0} \epsilon ).
\end{align}
We note that in the case
\begin{equation}\label{condi}
\begin{split}
&\epsilon < (2G_{0} M_{0})^{-1},
\\
&\frac{1}{L^2} < -6 G_{0} M_{0} \epsilon ^3 \ln (2 G_{0} M_{0} \epsilon ),
\end{split}
\end{equation}
this effective cosmological constant becomes a negative quantity and the solution turns into asymptotically de Sitter space.
Even more, in the case
\begin{equation}\label{condicero}
\begin{split}
&\epsilon < (2G_{0} M_{0})^{-1},
\\
&\frac{1}{L^2} = -6 G_{0} M_{0} \epsilon ^3 \ln (2 G_{0} M_{0} \epsilon ),
\end{split}
\end{equation}
the effective cosmological constant vanishes and the asymptotic (anti-) de Sitter behaviour disappears.
In other words,
the running parameter could be the responsible of certain topology change in the solution. A similar result can be
found in Ref. \cite{Koch:2014joab}, where an effective cosmological constant appears when the asymptotic behaviour of certain 
scale--dependent spherically symmetric space--time with cosmological coupling is considered.

\begin{figure}[ht!]
\centering
\includegraphics[width=\linewidth]{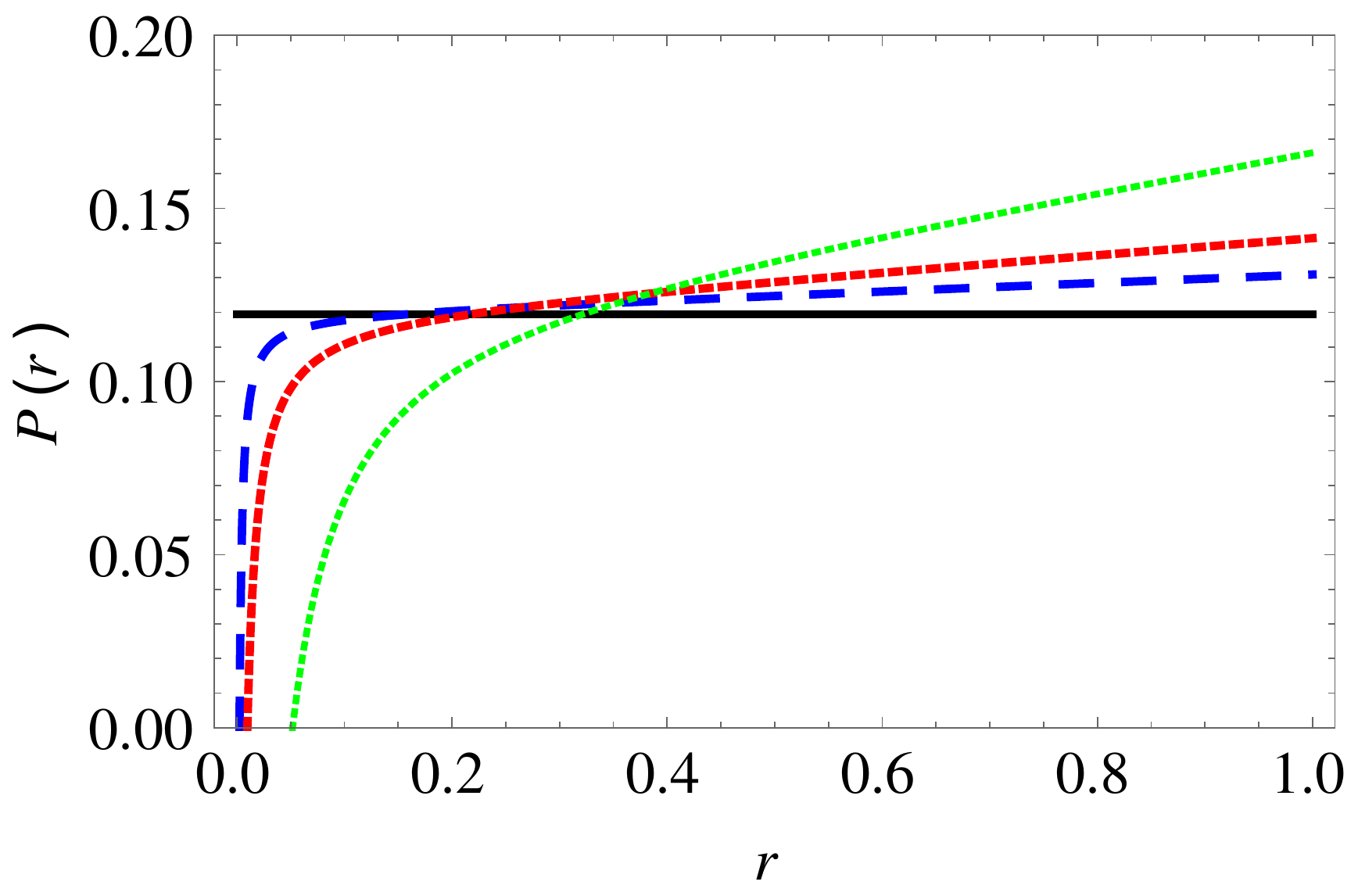}
\caption{\label{polyfig} Polytropic pressure function for
$\epsilon=0.00$ (black solid line), $\epsilon=0.05$ (dashed blue line), $\epsilon=0.10$ 
(short dashed red line), $\epsilon=0.25$ (dotted green line). The other values have been taken as unity.
}
\end{figure}

The behaviour of the scale--dependent polytropic pressure is shown in figure \ref{polyfig}. 
It can be seen that, depending on $\epsilon$, $P(r)$ diverges at  $r\to 0$ to $-\infty$ and 
behaves as 
\begin{eqnarray}\label{asinfp}
P(r)\approx\frac{3 (2 r \epsilon +1) }{8 \pi  G_{0}}\left[\frac{1}{L^2}+6 G_{0} M_{0} \epsilon ^3 \ln (2 G_{0} M_{0} \epsilon )\right],
\end{eqnarray}
in the limit $r\to \infty$. Even more, the asymptotic value of $P(r)$ in Eq. (\ref{asinfp}) can be a negative value whenever
\begin{align}
\begin{split}
&\epsilon < (2G_{0} M_{0})^{-1},
\\
&\frac{1}{L^2} < -6 G_{0} M_{0} \epsilon ^3 \ln (2 G_{0} M_{0} \epsilon ),
\end{split}
\end{align}
which coincides with the condition demanded for a de Sitter asymptotic behaviour of $f(r)$ in Eq. (\ref{condi}).

\begin{figure}[ht!]
\centering
\includegraphics[width=\linewidth]{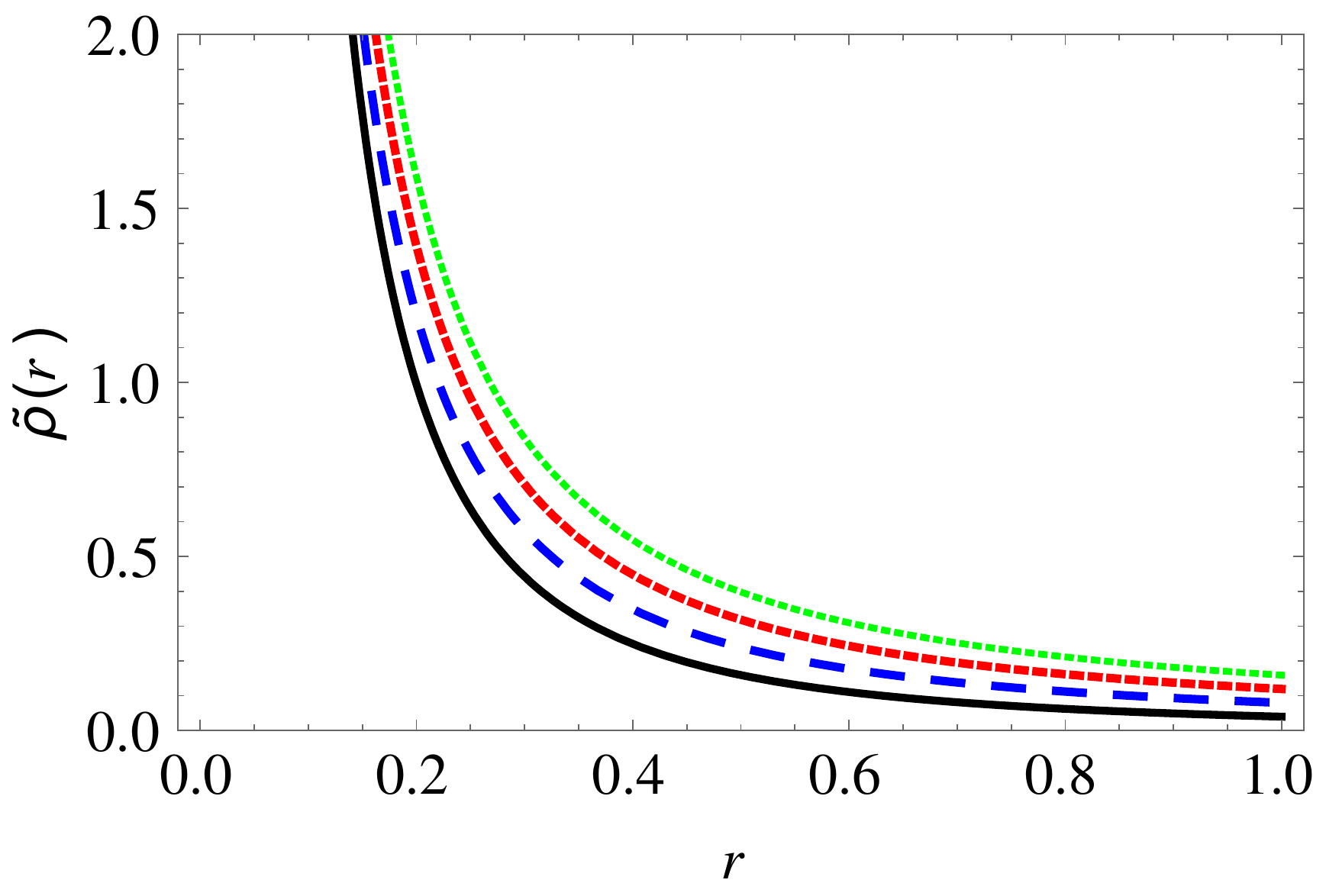}
\caption{\label{densifig} Scale--dependent density for $\epsilon=0.00$ (black solid line), $\epsilon=1.00$ (dashed blue line), $\epsilon=2.00$
(short dashed red line), $\epsilon=3.00$ (dotted green line). The other values have been taken as unity. See text for details.
}
\end{figure}

Figure \ref{densifig} shows the scale--dependent density profile $\tilde{\rho}$ for different values of $\epsilon$.
It can be seen that $\tilde{\rho}$ increases with the running parameter $\epsilon$. More precisely, $\tilde{\rho}$ decreases slowly compared with
the classical density. In fact, when $r\to\infty$, $\tilde{\rho}\sim r^{-1}$. 
Regarding the matter content, we remark that $(T^{\text{M}})_{\mu\nu}$ 
fullfils all the energy conditions as in the classical case.

We end this section noticing that the singularity at $r=0$ cannot be removed in the scale--dependent context. In fact, for
small $\epsilon$, the curvature scalars are given by
\begin{align}
&R \approx R_{0}-\frac{6  G_{0} M_{0}}{r^2}\epsilon, 
\\
&Ricc \approx Ricc_{0}+\frac{12 G_{0} M_{0}}{r^4} \left(\frac{3 r^2}{L^2}-1\right) \epsilon,
\\
&\mathcal{K} \approx \mathcal{K}_{0}-\frac{48 G_{0}^{2} M_{0}^{2}}{ r^5}\epsilon\ 
-\frac{24 G_{0} M_{0}}{ r^4}\left(1-\frac{r^2}{L^2}\right)\epsilon,
\end{align}
showing that the classical results are recovered in the limit $\epsilon\to 0$ and that the singularity persists at $r=0$. Furthermore,
$R\sim r^{-2}$, $Ricc\sim r^{-4}$ and $\mathcal{K}\sim r^{-6}$ for $r\to0$, in agreement with the classical case.

\subsection{Horizons and black hole Thermodynamics}
\begin{figure}[ht!]
\centering
\includegraphics[width=\linewidth]{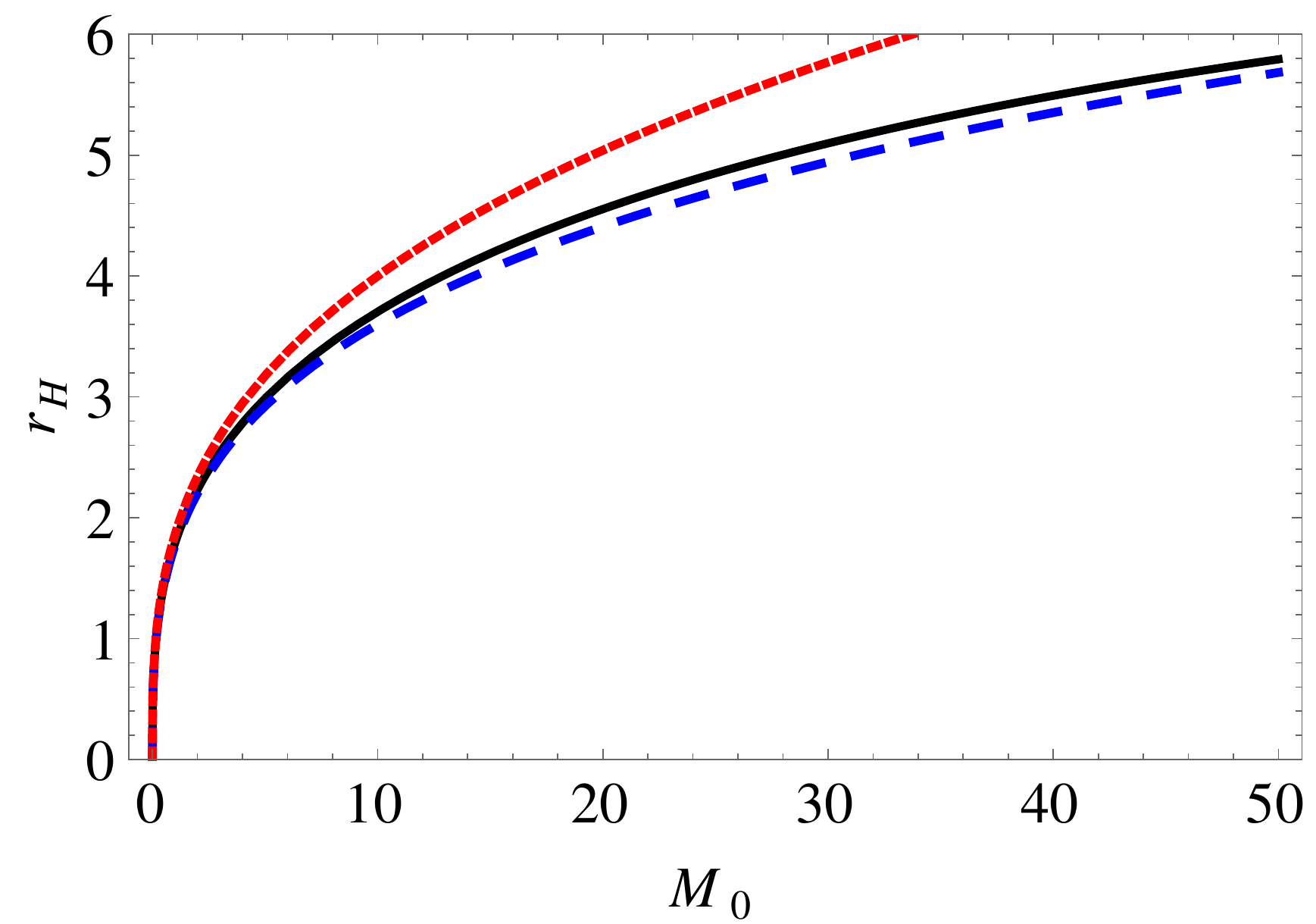}
\caption{\label{horizon} Horizon radius as a function of $M_{0}$ for
$\epsilon=0.05$, $L=2$ and $G_{0}=0.8$.  The plots for the numerical solution, the 
expansion for $\epsilon \ll 1$ and the classical case correspond to the
black solid line, dashed blue line and 
short dashed red line respectively.
}
\end{figure}

In order to study the thermodynamics of the scale--dependent polytropic BH, the
horizon radius, $r_{H}$, must be computed. As it is well known, $r_{H}$ can be obtained as one of the real positive roots of the lapse
function, $f(r)$. However, demanding $f(r_H)=0$ in our solution, leads to a transcendental equation for $r$ which must
be solved numerically. Besides, we can solve for $\epsilon \ll 1$ to obtain 
\begin{eqnarray} \label{r_H_Approx}
r_{H} \approx \left[\frac{(\sqrt{2 r_{0}^3 \epsilon ^3+4}+2)^{2/3}-2^{1/3}
r_{0} \epsilon}{(4(\sqrt{2 r_{0}^3 \epsilon ^3+4}+2))^{1/3}} \right]r_0 ,
\end{eqnarray}
where $r_{0}=\sqrt[3]{2L^{2}G_{0}M_{0}}$ is the classical horizon defined in Sect. \ref{poly_BH_section}.
In particular, if we take the limit of Eq. (\ref{r_H_Approx}) when $\epsilon \rightarrow 0$ one obtains that $r_H = r_0$, as it should be.
It can be seen that, for small $M_{0}$, the horizon radius obtained by both methods coincides with the classical one (see figure \ref{horizon}). 
Moreover, the scale--dependent $r_{H}$ deviates
from the classical horizon as $M_{0}$ increases.
In order to get more insight into the thermodynamics of the scale--dependent polytropic BH we
shall calculate the Hawking temperature, $T_{H}$, and the Bekenstein--Hawking entropy, $S_{BH}$. It is remarkably that,
although $r_{H}$ cannot be obtained analitically, we can
obtain $T_{H}$  and $S_{BH}$ as implicit functions of the horizon radius. In fact,
\begin{align}
&T_{H} = \frac{3 G_{0} M_{0}}{2\pi  r_{H}^2 (\epsilon r_{H}  +1 )},
\end{align}
\begin{align}
&S_{BH} = \frac{\mathcal{A}_{H}}{4G(r_H)} = S_0 \cdot (1 + \epsilon r_H ),
\end{align}
as can be checked by the reader. However, the behaviour of these thermodynamics quantities as a function of the classical
BH mass $M_{0}$, must be obtained after a numerical analysis is performed.

\begin{figure}[ht!]
\centering
\includegraphics[width=\linewidth]{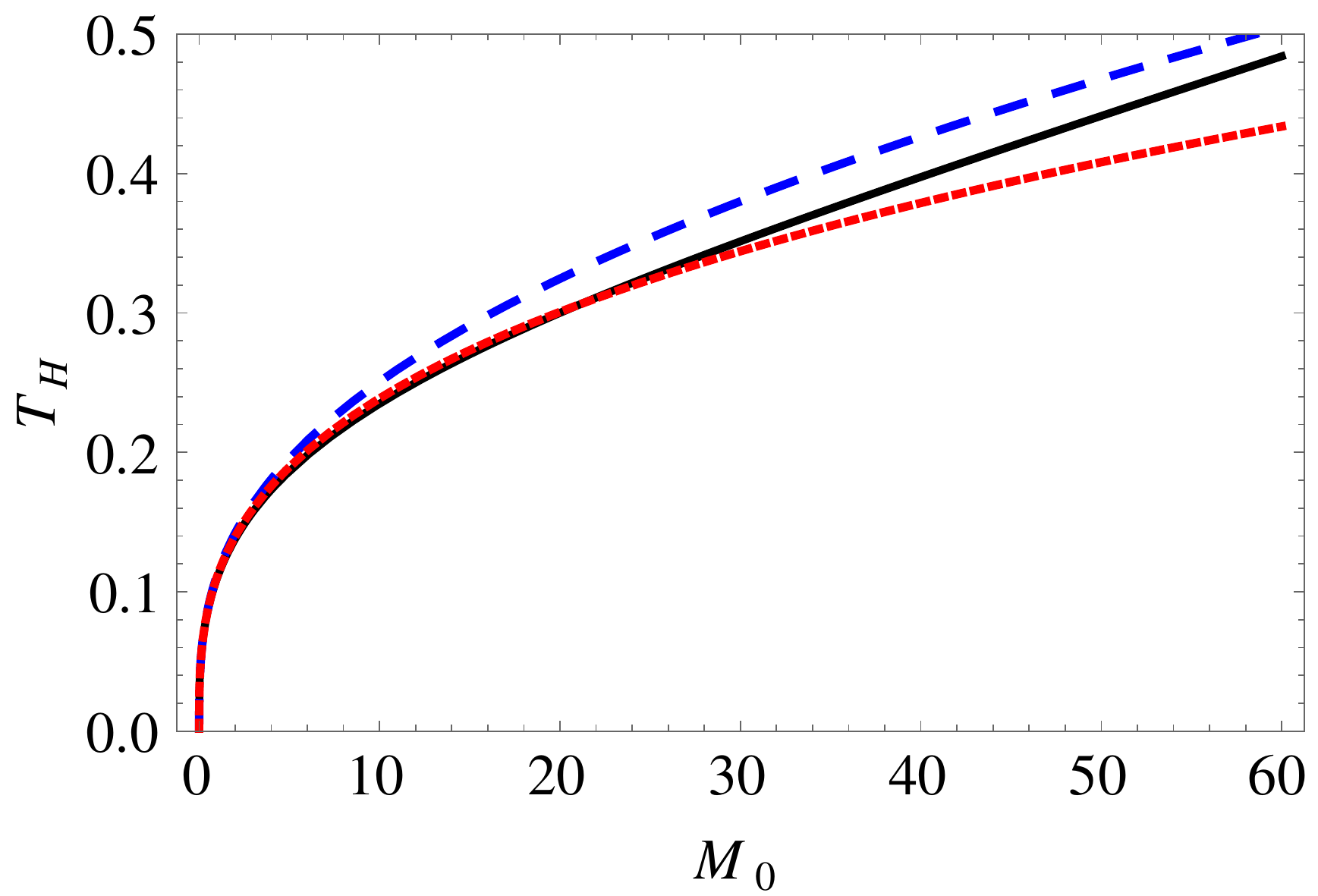}
\caption{\label{temp} Hawking temperature as a function of $M_{0}$ for
$\epsilon=0.05$, $L=2$ and $G_{0}=0.8$.  The plots for the numerical solution, the 
expansion for $\epsilon \ll 1$ and the classical case correspond to the
black solid line, dashed blue line and 
short dashed red line respectively.
}
\end{figure}

In figure \ref{temp} we compare the behaviour of $T_{H}$ obtained numerically
as a function of $M_{0}$ with the classical result and the one obtained for $\epsilon \ll 1$ which reads
\begin{eqnarray} \label{Tfinal}
T_{H}&\approx & T_0 + \frac{15 r_{0}^3 \epsilon ^2}{16 \pi  L^2}.
\end{eqnarray}
As in the previous case, the temperatures coincide for small classical BH mass and a deviation is observed as 
$M_{0}$ increases. 

\begin{figure}[ht!]
\centering
\includegraphics[width=\linewidth]{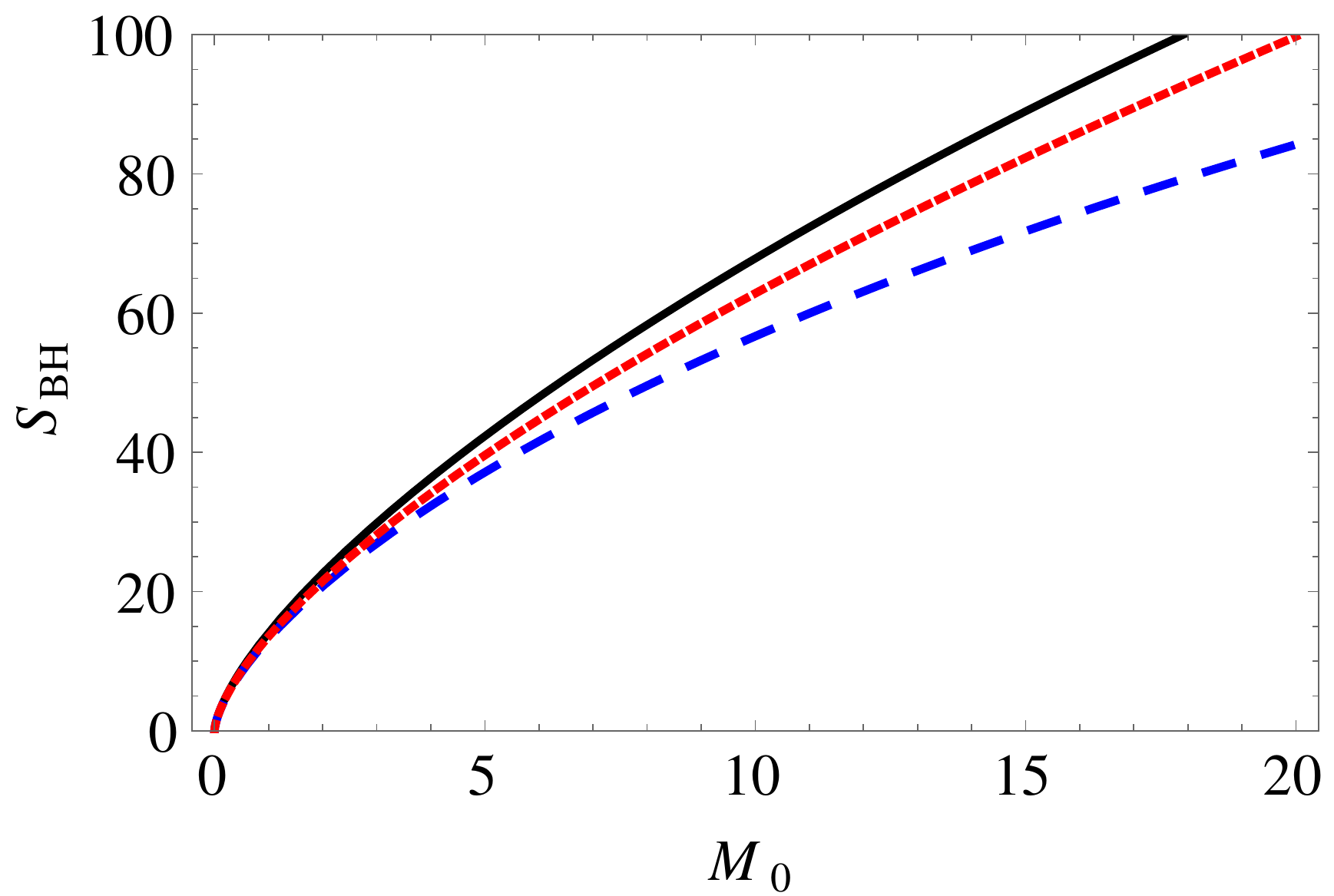}
\caption{\label{entro} Bekenstein-- Hawking entropy as a function of $M_{0}$ for
$\epsilon=0.07$, $L=2$ and $G_{0}=0.8$. The plots for the numerical solution, the 
expansion for $\epsilon \ll 1$ and the classical case correspond to the
black solid line, dashed blue line and 
short dashed red line, respectively.
}
\end{figure}

A similar analysis is shown in figure \ref{entro} for the entropy. 
In this case, the behaviour of $S_{BH}$ obtained numerically
as a function of $M_{0}$ is compared with the classical result and the obtained for $\epsilon \ll 1$ which reads
\begin{eqnarray}\label{Sfinal}
S_{BH}&\approx&S_{0}-\frac{5}{2} \pi   L^{2} M_{0}^{3} r_{0}^3\epsilon ^2,
\end{eqnarray}
as figure \ref{entro} shows.

As a final comment, we note that the classical behaviour of both the temperature and the entropy is not considerably affected
by the runnig. In this sense, in spite the metric changes with small values of the runnig (including topology changes),
the thermodynamics remains robust because the horizon remains close to the classical one.

\section{Concluding remarks}\label{remarks}

In  this  article,  we  have  studied  
for the first time
the  scale--dependence  of a polytropic black hole in a spherically symmetric four-dimensional space--time assuming a non-null cosmological coupling. After
presenting the model and the classical black hole solutions,
we have allowed for a scale--dependence of the gravitational as well as the cosmological coupling, and we have solved
the corresponding generalized field equations by imposing
the  null  energy  condition.  Besides, we have studied in detail the horizon  structure,  asymptopia and 
some thermodynamic properties.

As a mandatory remark, one should note that the scale--dependent approach introduces an effective contribution to the energy momentum 
tensor thought $\Delta t_{\mu \nu}$. Moreover, in agreement with the classical solution, the Schwarzschild ansatz is preserved. Regarding 
the event horizon, it is important to mention that it is not analytical and therefore it is not possible to get an explicit expression for it. However, we 
are able to obtain a closed formula for the temperature and the entropy, writing those quantities in terms of the horizon radius. Note that, for small 
values of $\epsilon$, the scale--dependent solutions are in agreement with the classical black hole, but
when $\epsilon$ take large values, a strong deviation appears. 

In addition, and following the philosophy of the scale--dependent scenario, this novel solution (including the thermodynamic 
properties) should be quite similar to the classical counterpart. This is because we expect that the incorporation of quantum 
corrections slightly modifies the usual behavior, which is in agreement with Eqs. \ref{r_H_Approx}, \ref{Tfinal}, \ref{Sfinal}.

As this is a general feature of several 
scale--dependent black hole solutions studied in the past \cite{Koch:2014joab,koch2016,Rincon:2017goj}, 
we conclude that black hole thermodynamics is robust against 
this kind of deformations of the gravitational theory.

To conclude, some final comments are in order. First, we would like to point out that all of the results 
obtained here are independent of the proportionality 
constant $K$ appearing in the polytropic equation of state. Even more, the choice $n=-1/3$ for the polytropic index 
and the particular fitting for the integration constants $C_{1}$ and $C_{2}$ in section \ref{pol_theory}, lead to solutions which are unaffected whether $K$ depends on a certain scale or not. It can be shown that, for a different choice of the polytropic index and integration constants the matter sector must be modified incorporating the scale--dependence on $K$. Second, we note that, although we have extended the polytropic solution given in \cite{setare2015}, the
same technique could be applied to different exact polytrope solutions with cosmological constant reported in
the past (see, for example \cite{Stuchlik2000,Bohmer2004}). Third, from the point of view of a possible astrophysical test of the
solution here presented, an important feature to be studied is the so--called static radius (frequently known as the 
turnaround radius) \cite{Stuchlik1999,Panotopoulos:2017clp}, which defines the 
equilibrium region between gravitational attraction and dark energy repulsion. Interestingly, this radius, computed for
several polytropic solutions, can be related to the maximum allowable size of a spherical cosmic structure as a function of its mass \cite{Stuchlik2011,Stuchlik2016,epjc2017,Panotopoulos:2017clp}. In this sense, astrophysical observations of large--scale structures could indirectly shed light on 
possible bounds on the running parameter (which would enter as a new ingredient of the static radius), which would indicate 
deviations from general relativity. 

These and other aspects, which are beyond the scope of the present work, are left for a future publication.
 
\section*{ACKNOWLEDGEMENTS}
The author E.C. would like to acknowledge Nelson Bolivar for fruitful discussion.
The author A.R. was supported by the  CONICYT-PCHA/Doctorado  Nacional/2015-21151658. 
The author B.K. was supported by the Fondecyt 1161150. 
The author P.B. was supported by the Faculty of Science and Vicerrector\'{\i}a de Investigaciones of Universidad de los Andes, 
Bogot\'a, Colombia.
P. B. dedicates this work to In\'es Bargue\~no--Dorta.

\end{document}